\documentclass[superscriptaddress,twocolumn,nofootinbib]{revtex4}

\usepackage{amsfonts}
\usepackage{graphicx}
\usepackage{dcolumn}
\usepackage{bm}
\usepackage[all]{xy}\CompileMatrices

\newcommand{\bea}{\begin{eqnarray}}
\newcommand{\eea}{\end{eqnarray}}
\newcommand{\beq}{\begin{equation}}
\newcommand{\eeq}{\end{equation}}
\newcommand{\del}{\partial}
\newcommand{\nn}{\nonumber}
\newcommand{\lishi}{\langle\!\langle}
\newcommand{\rishi}{\rangle\!\rangle}

\begin{document}

\title{Coulomb-gas formulation of $SU(2)$ branes and chiral blocks}
\author{Samuli Hemming}
 \email[]{samuli.hemming@helsinki.fi}
 \affiliation{%
Helsinki Institute of Physics, P.O.Box 64, University of Helsinki,
FIN-00014 Finland
}%
\author{Shinsuke Kawai}
 \email[]{shinsuke.kawai@helsinki.fi}
\affiliation{%
Helsinki Institute of Physics, P.O.Box 64, University of Helsinki,
FIN-00014 Finland
}%
\author{Esko Keski-Vakkuri}
 \email[]{esko.keski-vakkuri@helsinki.fi}
\affiliation{%
Helsinki Institute of Physics, P.O.Box 64, University of Helsinki,
FIN-00014 Finland
}%
\affiliation{%
Department of Physical Sciences, P.O.Box 64, University of Helsinki,
FIN-00014 Finland
}%
\date{\today}

\begin{abstract}
We construct boundary states in $SU(2)_k$ WZNW models using the bosonized Wakimoto free-field representation and study their properties.
We introduce a Fock space representation of Ishibashi states which are coherent states of bosons with zero-mode momenta (boundary Coulomb-gas charges) summed over certain lattices according to Fock space resolution of $SU(2)_k$.
The Virasoro invariance of the coherent states leads to families of boundary states including the B-type D-branes found by Maldacena, Moore and Seiberg, as well as the A-type corresponding to trivial current gluing conditions.
We then use the Coulomb-gas technique to compute exact correlation functions of WZNW primary fields on the disk topology with A- and B-type Cardy states on the boundary.
We check that the obtained chiral blocks for A-branes are solutions of the Knizhnik-Zamolodchikov equations. 
\end{abstract}

\preprint{HIP-2003-67/TH, hep-th/0403145}
\maketitle

\section{Introduction}

Conformal field theory (CFT) on a group manifold provides an
excellent test ground for the study of strings on curved
backgrounds. Such models are exactly solvable due to large
symmetries, while they are non-trivial enough to exhibit the
peculiarity of curved target spaces. In recent years, D-branes in
Wess-Zumino-Novikov-Witten (WZNW) models have been investigated by
many authors, and remarkable progress has been
made\cite{Pradisi:1995qy, Pradisi:1995pp, Klimcik:1997hp, Kato:1997nu, 
Recknagel:1998sb, Alekseev:1999bs, Stanciu:1999id, Felder:1999ka, Behrend:1999bn, 
Maldacena:2001ky, Petkova:2002yj, Gaberdiel:2002qa}.
In particular, it is now
understood that possible locations for stable branes are
quantized, and the branes wrap conjugacy
classes of the group manifold\cite{Alekseev:1998mc, Bachas:2000ik,
Pawelczyk:2000ah}.
 From the world-sheet point of view, branes are boundaries of a two
dimensional manifold. The world-sheet description is then a boundary CFT,
where the branes correspond to specific states, called boundary
states.
The CFT analysis of D-branes is based on the algebraic formulation of
boundary states for rational conformal theories, dating back
to the work of Cardy and Ishibashi\cite{Cardy:1989ir,
Ishibashi:1989kg}. For rational models, operator product expansions
involving boundary and various duality relations among the
coupling constants were studied in \cite{Cardy:1991tv,
Lewellen:1992tb, Pradisi:1996yd}. Hence, at least for simple
rational CFT models, using the algebraic consistency conditions
one can find which D-branes are allowed in a given closed-string
background. It is an important open problem to generalize such
analysis to
more complicated theories, including non-rational models.

In this paper, we reconsider D-branes on WZNW
backgrounds from a somewhat different perspective.
CFT can also be formulated using the free-field (Coulomb-gas)
representation\cite{FeiginFuchs, Dotsenko:1984nm,
Dotsenko:1985ad}, which is in a sense complementary to the
algebraic approach. The free-field approach 
allows one to compute exact correlation
functions in a constructive manner -- assembling pieces and
finding integral expressions -- without having to solve
complicated Knizhnik - Zamolodchikov equations\cite{Knizhnik:1984nr}.
Instead, one needs to work out the cohomology of the screen charges. 
The Coulomb-gas formalism has recently been extended to include boundary states,
in the case of Virasoro minimal models \cite{Kawai:2002vd, Kawai:2002pz} and
in the case of CFTs with W-algebra symmetries
\cite{Caldeira:2003zz}. The present paper
deals with a free-field formulation of boundary states on a WZNW
background.  This should provide a concrete representation of
D-branes and should also be useful for computing exact correlation
functions. Indeed, such a free-field description of branes for the
$U(1)_k$ WNZW model is well known\cite{Affleck:2000ws, Fuchs:1998fu, Maldacena:2001ky}. The
$U(1)_k$ model corresponds to a free boson compactified on a circle
of radius $R=\sqrt{2k}$ (where $k$ is a positive integer),
allowing an extended symmetry so that the
CFT becomes rational. The primary fields are then indexed by an integer $r$ which
is defined modulo $2k$. The boundary state
construction starts from finding a basis,
the Ishibashi states. They are coherent states of a bosonic field,
with the ground state momenta summed over certain lattices.
Corresponding to the $2k$ primary fields of $U(1)_k$,
there are \cite{Fuchs:1998fu} $2k$ A-type Ishibashi
states
\beq 
\vert A; r\rishi_{U(1)}=\prod_{n>0}e^{\frac 1n
a_{-n}\bar a_{-n}}\sum_{\ell\in {\mathbb Z}} \left\vert
\frac{r+2k\ell}{\sqrt{2k}},\frac{r+2k\ell}{\sqrt{2k}}\right\rangle,
\label{eqn:U1AIshi}
\eeq
where $a_n$ and $\bar a_n$ are
holomorphic and anti-holomorphic Heisenberg operators, the ground
states $\vert *,*\rangle$ are parameterized by the left- and
right-moving momenta (holomorphic and anti-holomorphic boundary
charges). Our convention is $r=0, \cdots, 2k-1$.
By a T-duality transformation along the circle, the A-type
states are related to $2k$ B-type Ishibashi states
\beq 
\vert B; r\rishi_{U(1)}=\prod_{n>0}e^{-\frac 1n a_{-n}\bar
a_{-n}}\sum_{\ell\in {\mathbb Z}} \left\vert
\frac{r+2k\ell}{\sqrt{2k}},-\frac{r+2k\ell}{\sqrt{2k}}\right\rangle.
\label{eqn:U1BIshi}
\eeq
From the $2k$ A-type Ishibashi states one can construct $2k$ A-type
Cardy states, 
\beq 
\vert\widetilde{A; r}\rangle_{U(1)}
=(2k)^{-\frac 14}\sum_{s=0}^{2k-1}e^{-\pi i rs/k}
\vert A; s\rishi_{U(1)}, 
\eeq 
which are interpreted as D0 branes at $2k$
special points on the circle. The B-type Cardy states are linear
combinations of only two of the $2k$ B-type Ishibashi states, 
\beq
\vert\widetilde{B; \eta}\rangle_{U(1)}=\left(\frac
k2\right)^{\frac 14}\left(\vert B; 0\rishi_{U(1)}+\eta\vert B;
k\rishi_{U(1)}\right), 
\eeq 
where $\eta=\pm 1$. The other $2k-2$
B-type Ishibashi states decouple from the theory\footnotemark[1].
\footnotetext[1]{In the standard language the A-type states
defined above are Dirichlet and the B-type states are Neumann.
It is however more natural to define them according to their coupling
to bulk operators, and then the distinction of A- and B-branes depends
on the convention of how we define the non-chiral (left$\times$ right)
bulk operators. This point is discussed in Section 3.} These B-type
Cardy states are interpreted as D1-branes with a Wilson line
parameterized by $\eta$. The main goal of this paper is to
construct a similar free-field representation of the $SU(2)_k$
branes by  finding a Fock space representation of the boundary
states for the parafermion part of $SU(2)_k$. In contrast to the
$U(1)_k$ part where the momenta are made periodic by the infinite
sum over the identified values in (\ref{eqn:U1AIshi}) and (\ref{eqn:U1BIshi}),
the momentum summation for the parafermion part is much more
complicated. This is because the sum over momenta becomes a lattice
sum with a non-trivial truncation of null vectors.
Such null vector structures must be examined
carefully in order that the boundary states satisfy the Cardy
condition. We argue that for the $SU(2)_k$ model the momenta
should be summed over a lattice which is basically a Kaluza-Klein
tower but modified according to the structure of null vectors; in
the Coulomb-gas terminology such a lattice sum is over the genus
one Felder complex arising from the Fock space resolution of the
spectra on the torus.

The idea of realizing WZNW branes in the
Wakimoto free field representation is not new. The A-branes of the
$SU(2)_k$ model are constructed using the standard Wakimoto
representation in \cite{Ishikawa:2000ez} (see also \cite{Deliduman:2002bf})
and their BRST property
is discussed in \cite{Parkhomenko:2001ki}. The novelty of this
paper is a more general treatment, which naturally includes the
B-branes of $SU(2)_k$ found in \cite{Maldacena:2001ky}
(see \cite{Fuchs:1997kt, Fuchs:1998fu, Birke:1999ik, Fuchs:1999zi, Fuchs:1999xn}
for general discussions), 
and the
computational method to obtain exact correlation functions of bulk
primary operators using the Coulomb-gas screening charges. For
these purposes we shall employ a bosonic version of the Wakimoto
construction, which renders the symmetry of the model more
transparent.

The free-field realization of branes gives another technical tool for calculations
-- sometimes a more convenient one than the other approaches. 
For example, in the Coulomb-gas formalism correlation functions involving such 
branes may in principle be generalized to arbitrary topology by sewing smaller 
diagrams together; 
on the other hand solving Knizhnik-Zamolodchikov equations for higher 
topologies is notoriously difficult.

The plan of this paper is as follows. In the next section we
review the bosonized Wakimoto free-field representation of WZNW
models and collect necessary ingredients for the following
discussions. In Section 3 we construct boundary states of the
$SU(2)_k$ and the parafermion $SU(2)_k/U(1)_k$ models using the
free-field representation. We present sample computations of
correlation functions involving the boundary states in Section 4.
We summarize the results and conclude in Section 5.


\section{Bosonic Wakimoto representation of $SU(2)_k$ WNZW model}
In this section we review the Coulomb-gas representation of the $SU(2)_k$ WNZW model\cite{Dotsenko:1990ui, Dotsenko:1991zb, Distler:1990xv, Konno:1992vm, Itoh:1993mt, Jayaraman:1990tu, Nemeschansky:1989wg, Nemeschansky:1991rx, Bilal:1989py, Frau:1990pt, Gerasimov:1990fi}.
We follow the bosonic construction of \cite{Jayaraman:1990tu, Nemeschansky:1989wg, Nemeschansky:1991rx, Bilal:1989py, Frau:1990pt, Gerasimov:1990fi} which emphasizes the fact that $SU(2)_k$ can be written as a product of ${\mathbb Z}_k$ parafermions and a $U(1)_k$ boson.
We shall spell out anti-holomorphic expressions explicitly when they differ from the holomorphic counterparts, as they will be necessary for discussing boundary states.


\subsection{Wakimoto free field representation and bosonization}


We start from the Gauss decomposition of the group element $g$ in $SL(2,{\mathbb C})$,
\beq
g=
\left(\begin{array}{cc}
1 & \bar\gamma \\ 0& 1
\end{array}\right)
\left(\begin{array}{cc}
e^{\tilde\varphi} & 0 \\ 0& e^{-\tilde\varphi}
\end{array}\right)
\left(\begin{array}{cc}
1 & 0 \\ \gamma & 1
\end{array}\right).
\eeq
This leads to the Wakimoto representation of the $SU(2)$ currents
$J=kg^{-1}\del g$ and $\bar J=-k(\bar\del g)g^{-1}$, that is\cite{Gerasimov:1990fi, Ishikawa:2000ez},
\bea
J^+&=&\beta,\\
J^-&=&i\sqrt{2(k+2)}\del\varphi\gamma-k\del\gamma-\beta\gamma^2,\\
J^3&=&-i\sqrt{\frac{k+2}{2}}\del\varphi+\gamma\beta,
\eea
and
\bea
\bar J^+&=&-i\sqrt{2(k+2)}\bar\del\bar\varphi\bar\gamma+k\bar\del\bar\gamma+\bar\beta\bar\gamma^2,\\
\bar J^-&=&-\bar\beta,\\
\bar J^3&=&i\sqrt{\frac{k+2}{2}}\bar\del\bar\varphi-\bar\gamma\bar\beta,
\eea
where the usual right-nested normal ordering is implicit.
The fields $\beta$ and $\bar\beta$ have been introduced as auxiliary fields. 
$\tilde\varphi$ has been rescaled and $\varphi$, $\bar\varphi$ are respectively its 
holomorphic and anti-holomorphic part.
These currents and Sugawara stress tensor generate the $SU(2)$ affine Lie algebra,
\bea
T(z)T(z')&\sim&\frac{c/2}{(z-z')^4}+\frac{2T(z')}{(z-z')^2}+\frac{\del T(z')}{(z-z')},\\
T(z)J^a(z')&\sim&\frac{J^a(z')}{(z-z')^2}+\frac{\del J^a(z')}{(z-z')},\\
J^a(z)J^b(z')&\sim&\frac{(k/2)\delta^{ab}}{(z-z')^2}+\frac{if^{ab}{}_{c}}{(z-z')}J^c(z'),
\eea
and likewise for the anti-holomorphic part,
with central charge
\beq
c=\frac{3k}{k+2}
\eeq
and structure constants $f^{12}{}_3=f^{23}{}_1=f^{31}{}_2=1$.
The Wakimoto free-fields have operator products
\bea
&&\varphi(z)\varphi(z')\sim-\ln(z-z'),\nn\\
&&\beta(z)\gamma(z')\sim\frac{-1}{z-z'},\;\;\gamma(z)\beta(z')\sim\frac{1}{z-z'},
\eea
and similarly for $\bar\varphi$, $\bar\beta$, and $\bar\gamma$.
We bosonize the $\beta$-$\gamma$ system,
\bea
\beta&=&-i\del\chi e^{\eta-i\chi},\\
\gamma&=&e^{-\eta+i\chi},
\eea
and redefine the fields by a linear transformation,
\bea
\phi^{(1)}&=&\varphi-i\sqrt{\frac{k+2}{2}}\eta-\sqrt{\frac{k+2}{2}}\chi,\nn\\
\phi^{(2)}&=&i\sqrt{\frac{k+2}{k}}\varphi+\frac{k+2}{\sqrt{2k}}\eta-i\sqrt{\frac k2}\chi,\nn\\
\phi^{(3)}&=&-\sqrt{\frac{k+2}{k}}\varphi+i\sqrt{\frac 2k}\eta.
\label{eqn:lintransf1}
\eea
Our convention for the operator products between the new bosonic
fields is
\beq
\phi^{(i)}(z)\phi^{(j)}(z')\sim-\delta^{ij}\ln(z-z'),
\eeq
where the indices $i$ and $j$ run 1, 2, 3.
After the redefinition,
the holomorphic parts of the chiral current take the form
\bea
J^+(z)&=&-\sqrt k\Psi(z)e^{i\sqrt{\frac 2k}\phi^{(3)}},\nn\\
J^-(z)&=&-\sqrt k\Psi^\dag(z)e^{-i\sqrt{\frac 2k}\phi^{(3)}},\nn\\
J^3(z)&=&i\sqrt\frac k2\del\phi^{(3)},
\label{eqn:J}
\eea
where $\Psi(z)$ and $\Psi^\dag(z)$ are the parafermion currents,
\bea
\Psi(z)&=&\frac{1}{\sqrt 2}\left(-i\sqrt{\frac{k+2}{k}}\del\phi^{(1)}+\del\phi^{(2)}\right)
e^{\sqrt{\frac 2k}\phi^{(2)}},\nn\\
\Psi^\dag(z)&=&\frac{1}{\sqrt 2}\left(-i\sqrt{\frac{k+2}{k}}\del\phi^{(1)}-\del\phi^{(2)}\right)
e^{-\sqrt{\frac 2k}\phi^{(2)}}.\nn\\
\eea
Thus the two bosons $\phi^{(1)}$ and $\phi^{(2)}$ constitute
the bosonic representation of the ${\mathbb Z}_k$ parafermions,
whereas the $U(1)_k$ part is represented by $\phi^{(3)}$.

The stress tensor takes the form
\bea
T(z)
&=&-\frac 12 \delta_{ij}\del\phi^{(i)}\del\phi^{(j)}+2i\alpha_0\del^2\phi^{(1)}\nn\\
&=&-\frac 12 \del\phi\cdot\del\phi+2i\alpha_0\rho\cdot\del^2\phi,
\label{eqn:stresstensor}
\eea
where
\beq
\alpha_0=\frac{1}{2\sqrt{2(k+2)}},
\label{eqn:alpha0andk}
\eeq
and $\rho=(1,0,0)$.

Bosonization of the anti-holomorphic fields,
\bea
\bar\beta&=&i\bar\del\bar\chi e^{\bar\eta+i\bar\chi},\\
\bar\gamma&=&e^{-\bar\eta-i\bar\chi},
\eea
and redefinition of the fields
\bea
\bar\phi^{(1)}&=&\bar\varphi-i\sqrt{\frac{k+2}{2}}\bar\eta+\sqrt{\frac{k+2}{2}}\bar\chi,\nn\\
\bar\phi^{(2)}&=&i\sqrt{\frac{k+2}{k}}\bar\varphi+\frac{k+2}{\sqrt{2k}}\bar\eta+i\sqrt{\frac k2}\bar\chi,\nn\\
\bar\phi^{(3)}&=&-\sqrt{\frac{k+2}{k}}\bar\varphi+i\sqrt{\frac 2k}\bar\eta,
\label{eqn:lintransf2}
\eea
lead to bosonic expressions of the anti-holomorphic currents and the stress tensor,
\bea
&&\bar J^+(\bar z)\nn\\
&&\;\;\;=\left(-i\sqrt{\frac{k+2}{2}}\bar\del\bar\phi^{(1)}-\sqrt{\frac k2}\bar\del\bar\phi^{(2)}\right)
e^{-\sqrt{\frac 2k}(\bar\phi^{(2)}+i\bar\phi^{(3)})},\nn\\
&&\bar J^-(\bar z)\nn\\
&&\;\;\;=\left(-i\sqrt{\frac{k+2}{2}}\bar\del\bar\phi^{(1)}+\sqrt{\frac k2}\bar\del\bar\phi^{(2)}\right)
e^{\sqrt{\frac 2k}(\bar\phi^{(2)}+i\bar\phi^{(3)})},\nn\\
&&\bar J^3(\bar z)=-i\sqrt\frac k2\bar\del\bar\phi^{(3)},\label{eqn:Jbar}\\
&&\bar T(\bar z)
=-\frac 12 \bar\del\bar\phi\cdot\bar\del\bar\phi+2i\alpha_0\rho\cdot\bar\del^2\bar\phi.
\label{eqn:Tbar}
\eea

From the form of the stress tensor (\ref{eqn:Tbar}), it is obvious that flipping and rotating the bosonic fields in the directions of $\bar\phi^{(2)}$ and $\bar\phi^{(3)}$ leave the stress tensor invariant.
To be more specific, we define transformations $\omega$ of the anti-holomorphic bosonic fields $\bar\phi^{(i)}$ which operate trivially on $\bar\phi^{(1)}$ but linearly transform the remaining two components,
\beq
\omega:
\left(\begin{array}{c}
\bar\phi^{(1)}\\ \bar\phi^{(2)}\\ \bar\phi^{(3)}\end{array}\right)
\mapsto
\left(\begin{array}{cc}
1&\begin{array}{cc}0&0\end{array}\\
\begin{array}{c}0\\0\end{array}&\tilde M
\end{array}\right)
\left(\begin{array}{c}\bar\phi^{(1)}\\\bar\phi^{(2)}\\\bar\phi^{(3)}\end{array}\right),
\label{eqn:omega}
\eeq
where $\tilde M$ is a $2\times 2$ matrix.
The invariance of $\bar T(\bar z)$ implies that
$\tilde M$ must be orthogonal\footnotemark[2], $\tilde M^T \tilde M=1$.
\footnotetext[2]{
It may be more natural to consider $\omega$ as {\em antilinear} transformations and $\tilde M$ (and also $M$ in Section 3) being (anti-)unitary, so that the rotation between $\bar\phi^{(2)}$ and $\bar\phi^{(3)}$ is interpreted as $SU(2)$ rotation within conjugacy classes.
This should lead to a different choice of basis in the Hilbert space of the anti-holomorphic sector (see section 5).}
Under the transformations $\omega$, the anti-holomorphic
$SU(2)$ currents are not invariant.
We shall denote the transformed currents as
\beq
\Omega \bar J^a(\bar z)=\bar J^a(\omega\bar\phi^{(i)}(\bar z)) \ .
\eeq
That is, $\Omega\bar J^a(\bar z)$ are the currents constructed as (\ref{eqn:Jbar}) but
with the transformed bosonic fields $\omega\bar\phi^{(i)}$ instead of $\bar\phi^{(i)}$.
Note that
\beq
\Omega\bar J^\pm=\bar J^\pm \ , \ \Omega\bar J^3=\bar J^3
\eeq
for $\tilde M={\rm diag}(1, 1)$ and
\beq
\Omega\bar J^\pm=\bar J^\mp \ , \ \Omega\bar J^3=-\bar J^3
\eeq
for $\tilde M={\rm diag}(-1, -1)$. In general,
$\Omega \bar J^a$ cannot be written in a simple form.

The chiral $SU(2)$ primary fields $\Phi_{j,m}(z)$ with isospin $j$ and
magnetic quantum number $m$ are represented by the vertex operators
\beq
V_{j,m}(z)=K_{j,m} \exp\left(i\alpha_{j,m}\cdot\phi(z)\right),
\label{eqn:vertexop}
\eeq
where $K_{j,m}=[(2j)!/(j+m)!(j-m)!]^{1/2}$ and
\beq
\alpha_{j,m}=\left(-j\sqrt{\frac{2}{k+2}}, -im\sqrt{\frac 2k}, m\sqrt{\frac 2k}\right).
\label{eqn:alphajm}
\eeq
Using operator products with the stress tensor, their
conformal dimensions are verified to be
\beq
h_{j,m}=\frac 12 \alpha_{j,m}\cdot (\alpha_{j,m}-4\alpha_0\rho) = \frac{j(j+1)}{k+2}.
\eeq
The same primary field $\Phi_{j,m}(z)$ can also be represented by
another vertex operator,
\beq
V^\dag_{j,m}(z)=K_{j,m} \exp\left(i\alpha^\dag_{j,m}\phi(z)\right),
\eeq
where the conjugate charge $\alpha^\dag_{j,m}$ is
\bea
\alpha^\dag_{j,m}
&=&4\alpha_0\rho-\alpha_{j,-m}\nn\\
&=&\left((1+j)\sqrt{\frac{2}{k+2}}, -im\sqrt{\frac 2k}, m\sqrt{\frac 2k}\right).
\eea
The equivalence of $V_{j,m}(z)$ and $V^\dag_{j,m}(z)$ is a basic
feature of the Coulomb-gas formalism. It can be used to minimize
the number of screening operators required for non-vanishing
correlation functions.
In the $SU(2)_k$ model this is related to the equivalence of $\Phi_{j, m}(z)$ and $\Phi_{\frac k2 -j, \frac k2+m}(z)$ which is sometimes called spectral flow identification.

Non-chiral (left $\times$ right) vertex operators are simply
direct products of chiral operators, but since there are two equivalent
vertex operator representations for a single chiral field, there are
four ways to express a non-chiral primary field $\Phi_{j,m}(z,\bar
z)$:
\bea
&&V_{j,m}(z)\bar V_{j,m}(\bar z), \;\;
V_{j,m}(z)\bar V^\dag_{j,m}(\bar z), \nn\\
&&V^\dag_{j,m}(z)\bar V_{j,m}(\bar z), \;\;
V^\dag_{j,m}(z)\bar V^\dag_{j,m}(\bar z),
\label{eqn:su2vertex}
\eea
where the right moving part of the vertex operators are
\bea
\bar V_{j,m}(\bar z)&=&K_{j,m} \exp\left(i\alpha_{j,m}\cdot\bar\phi(\bar z)\right),\nn\\
\bar V^\dag_{j,m}(\bar z)&=&K_{j,m} \exp\left(i\alpha^\dag_{j,m}\cdot\bar\phi(\bar z)\right).
\eea


\subsection{Truncation of Fock modules and characters}


The key element of the free-field formalism is the screening operators
which control the structure of singular vectors.
They are used for finding integral expressions of
correlation functions, and are also used to construct Felder's BRST operators whereby
physical spectra of CFT are realized as their
cohomology spaces \cite{Felder:1989zp, Bernard:1990iy, Bouwknegt:1990jf, Bouwknegt:1990xa, Jayaraman:1990tu}.

We focus on the parafermion part of $SU(2)_k$, as the $U(1)_k$ part is trivial.
The parafermion primary fields $\Phi_{l,n}^{PF}(z)$ labelled by two integers
$l=0,1,\cdots,k$ and $n=-k,-k+1,\cdots,k-1$ are represented by vertex operators
\beq
V_{l,n}^{PF}(z)=K^{PF}_{l,n} \exp\left\{-\frac{il}{\sqrt{2(k+2)}}\phi^{(1)}+\frac{n}{\sqrt{2k}}\phi^{(2)}\right\},
\label{eqn:pfvertex}
\eeq
for $-l\leq n\leq l$.
The normalisation constant is $K^{PF}_{l,n}=[l!/((l+n)/2)!((l-n)/2)!]^{1/2}$.
Viewing the parafermions independently, as opposed to as a part of
$SU(2)_k$, there are also operators $\Phi_{l,n}^{PF}(z)$ with $n>l$.
They are obtained by successive application of $\Psi(z)$ on (\ref{eqn:pfvertex}).
Acting on the vacuum, the vertex operators $V^{PF}_{l,n}$
generate a state space, the Fock module $F_{l,n}$.

There are three screening operators,
\beq
Q_1=\oint dz V_1(z), \;\;
Q_\pm=\oint dz V_\pm (z),
\eeq
where
\bea
V_1(z)&=&\del\phi^{(2)}\exp\left\{i\sqrt{\frac{2}{k+2}}\phi^{(1)}\right\}, \\
V_\pm(z)&=&\exp\left\{-i\sqrt{\frac{k+2}{2}}\phi^{(1)}\pm\sqrt{\frac k2}\phi^{(2)}\right\}.
\eea
These screening operators satisfy the relations,
\bea
&& Q_+^2=Q_-^2=0, \nn\\
&& Q_+Q_-+(-1)^k Q_-Q_+=0, \nn\\
&& Q_\pm Q_1+Q_1Q_\pm=0.
\eea
The screening operators act on the Fock module $F_{l,n}$
as
\bea
Q_\pm F_{l,n}&=&F_{l+k+2,n\pm k}, \nn\\
Q_r F_{l,n}&=&F_{l-2r,n},
\eea
where
\beq
Q_r=\frac 1r \frac{e^{\frac{2\pi i r}{k+2}}-1}{e^{\frac{2\pi i}{k+2}}-1}
\oint dz_1\cdots \oint dz_r V_1(z_1)\cdots V_1(z_r),
\eeq
with the contours following Felder's convention \cite{Felder:1989zp}.
Repeated application of these operators on the Fock modules
generates an infinite diagram
${\cal C}_{PF}$ (figure 1).
\begin{figure}
\xymatrix{
F_{-l-2,n+2k} \ar@{.>}[dr] &F_{l,n+2k} \ar@{=>}[l]_{Q_{l+1}} \ar@{.>}[dr] &
F_{2k-l+2,n+2k} \ar@{=>}[l]_{\!\!\! Q_{k-l+1}} \ar@{.>}[dr] & {\cdots}\ar@{=>}[l]_{\;\;\;\; Q_{l+1}}\\
F_{l-k-2,n+k} \ar[ur] \ar@{.>}[dr] & F_{k-l,n+k} \ar@{=>}[l]_{Q_{k-l+1}} \ar[ur] \ar@{.>}[dr] &
F_{l+k+2,n+k} \ar@{=>}[l]_{Q_{l+1}} \ar[ur] \ar@{.>}[dr] & {\cdots} \ar@{=>}[l]_{\;\;\;\; Q_{k-l+1}}\\
F_{-l-2,n} \ar[ur] \ar@{.>}[dr] & F_{l,n} \ar@{=>}[l]_{Q_{l+1}} \ar[ur] \ar@{.>}[dr] &
F_{2k-l+2,n} \ar@{=>}[l]_{Q_{k-l+1}} \ar[ur] \ar@{.>}[dr] & {\cdots} \ar@{=>}[l]_{\;\;\;\; Q_{l+1}}\\
F_{l-k-2,n-k} \ar[ur] \ar@{.>}[dr] &F_{k-l,n-k} \ar@{=>}[l]_{Q_{k-l+1}} \ar[ur] \ar@{.>}[dr] &
F_{l+k+2,n-k} \ar@{=>}[l]_{Q_{l+1}} \ar[ur] \ar@{.>}[dr] & {\cdots} \ar@{=>}[l]_{\;\;\;\; Q_{k-l+1}}\\
F_{-l-2,n-2k} \ar[ur] & F_{l,n-2k} \ar@{=>}[l]_{Q_{l+1}} \ar[ur] &
F_{2k-l+2,n-2k} \ar@{=>}[l]_{Q_{k-l+1}} \ar[ur] & {\cdots} \ar@{=>}[l]_{\;\;\;\; Q_{l+1}}
}
\caption{
A part of the infinite diagram ${\cal C}_{PF}$ generated by $Q_{l+1}$, $Q_{k-l+1}$ and $Q_\pm$ for the parafermion module.
The solid and dotted arrows indicate the operations with $Q_+$ and $Q_-$, respectively.
}
\end{figure}
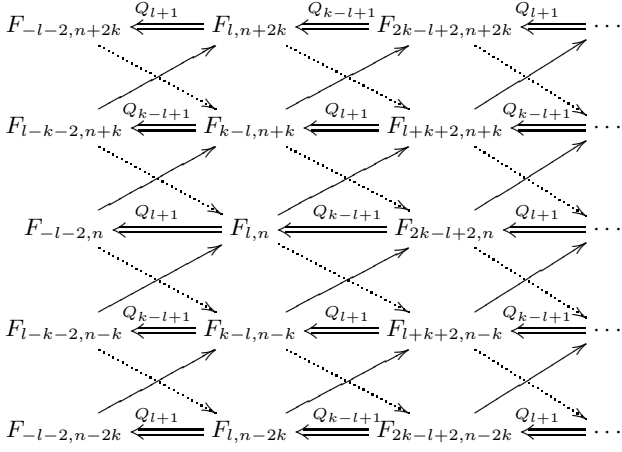
The physical spectrum in the parafermion
module is expected to be realized by the cohomology space,
\beq
\frac{{\rm Ker} Q_+ \cap {\rm Ker} Q_-}{{\rm Im} Q_+Q_-}\cap
\frac{{\rm Ker} Q_{l+1}}{{\rm Im} Q_{k-l+1} \cap {\rm Ker} Q_{l+1}} .
\eeq
The characters (string functions) of the parafermion modules
are calculated in \cite{Jayaraman:1990tu, Nemeschansky:1991rx,
Frau:1990pt}.
By assuming that the cohomology of
$({\rm Ker} Q_+\cap {\rm Ker Q_-})/ {\rm Im} Q_+Q_-$ is trivial on
the Fock module $F_{r,s}$ except for $(r,s)=(l,n)$, the characters are
found to be
\bea
\chi^{PF}_{l,n}(\tau)
&=&{\rm Tr}_{\frac{{\rm Ker} Q_+ \cap {\rm Ker} Q_-}{{\rm Im} Q_+Q_-}(F_{l,n})} q^{L_0-\frac{c}{24}}\nn\\
&=&\frac{1}{\eta(\tau)^2}\sum_{\stackrel
{\stackrel{\scriptstyle n_1, n_2 \in {\mathbb Z}/2}{\scriptstyle n_1-n_2\in {\mathbb Z}}}
{n_1\ge\vert n_2\vert, -n_1>\vert n_2\vert}}
(-1)^{2n_1}{\rm sign}(n_1)\nn\\
&&\times q^{\frac{(l+1+2(k+2)n_1)^2}{4(k+2)}-\frac{(n+2kn_2)^2}{4k}},
\label{eqn:pfchar}
\eea
where $q=\exp(2\pi i\tau)$.
The lattice sum is made for both $n_1\ge\vert n_2\vert$ and $n_1<-\vert n_2\vert$ wedges, and we define ${\rm sign}(0)=+1$.
This expression agrees with the parafermion character formula obtained by other means\cite{Kac:1984mq, Fateev:1985mm}.

The $SU(2)_k$ character for the isospin $j$ representation
is constructed from the parafermion characters as
\cite{Jayaraman:1990tu,Gepner:1987hr,Maldacena:2001ky}
\beq
\chi_j^{SU(2)}(\tau)
=\frac{1}{\eta(\tau)}\sum_n\sum_{n_3\in{\mathbb Z}}\chi^{PF}_{2j,n}q^{k(n_3+\frac{n}{2k})^2},
\label{eqn:su2char}
\eeq
The equivalence of this expression and the well known $SU(2)_k$ character formula,
\beq
\chi_j^{SU(2)}=\frac{\Theta_{2j+1,k+2}-\Theta_{-2j-1,k+2}}{\Theta_{1,2}-\Theta_{-1,2}},
\eeq
is shown for example in Appendix A of \cite{Jayaraman:1990tu}.
We use the Dedekind eta and Jacobi theta functions defined by
\bea
&&\eta(\tau)=q^{\frac{1}{24}}\prod_{n=1}^{\infty}(1-q^n),\\
&&\Theta_{\lambda, \mu}(\tau)=\sum_{n\in{\mathbb Z}}q^{(2\mu n+\lambda)^2/4\mu}.
\eea


\section{Fock space representation of boundary states}
The boundary states for D-branes in WZNW models have to satisfy
consistency conditions. One of them is the Ishibashi condition, i.e.
the conservation of energy momentum across the boundary. Second, they must
give appropriate overlaps for bulk chiral representations of the CFT.
In this section we find the
boundary states satisfying these conditions
for the $SU(2)_k$ model.
A nice feature of working in the free field Coulomb gas approach
is that we can start from a coherent state ansatz,
as in the case of string theory in a flat
background \cite{Callan:1987px, Polchinski:1988tu, Ishibashi:1989tf}
(for an introductory review, see e.g. \cite{DiVecchia:1999rh, DiVecchia:1999fx}).


\subsection{Construction of A-type $SU(2)_k$ boundary states}


We consider the (non-chiral) Fock spaces of the bosons $\phi^{(i)}$ and $\bar\phi^{(i)}$ discussed in the previous section.
The mode operators are defined through the expansion,
\beq
\phi^{(i)}(z)=\varphi_0^{(i)}-ia_0^{(i)}\ln z+i\sum_{n\neq 0}\frac{a_n^{(i)}}{n}z^{-n},
\eeq
where $z$ is a complex coordinate of the full plane.
We assume the standard radial quantization on the $z$-plane and the mode operators satisfy the Heisenberg algebra,
\bea
&&[a_m^{(i)}, a_n^{(j)}]=m\delta^{ij}\delta_{m+n,0},\nn\\
&&[\varphi_0^{(i)}, a_0^{(j)}]=i\delta^{ij}.
\eea
The anti-holomorphic counterpart $\bar\phi^{(i)}(\bar z)$ is expanded similarly and the mode operators satisfy the same algebra.
The holomorphic and anti-holomorphic Heisenberg operators are assumed to be independent.
In particular,
$[\varphi_0^{(i)}, \bar a_0^{(j)}] = [\bar\varphi_0^{(i)}, a_0^{(j)}] = 0$.
We denote the ground states of the Fock spaces as
$\vert\alpha,\bar\alpha;\alpha_0\rangle=\vert \alpha^{(i)}, \bar \alpha^{(i)}; \alpha_0\rangle$,
which are annihilated by the positive modes $a^{(i)}_{n>0}$, $\bar a^{(i)}_{n>0}$, and
$\alpha^{(i)}$ and $\bar\alpha^{(i)}$ are eigen values of the zero-mode momenta,
\bea
&&a_0^{(i)}\vert \alpha, \bar \alpha; \alpha_0\rangle
=\alpha^{(i)}\vert \alpha, \bar \alpha; \alpha_0\rangle,\nn\\
&&\bar a_0^{(i)}\vert \alpha, \bar \alpha; \alpha_0\rangle
=\bar \alpha^{(i)} \vert \alpha, \bar \alpha; \alpha_0\rangle.
\eea
These ground states are regarded as being constructed on the $SL(2,{\mathbb C})$ invariant vacuum $\vert 0,0;\alpha_0\rangle$ ($\alpha_0$ is to remind the existence of the non-trivial background charge) by applying  the vertex operators,
\bea
\vert \alpha,\bar\alpha; \alpha_0\rangle
&=&\lim_{z,\bar z\rightarrow 0} V_\alpha(z)\bar V_{\bar\alpha}(\bar z)\vert 0,0;\alpha_0\rangle\nn\\
&=&e^{i\alpha\cdot\varphi_0}e^{i\bar\alpha\cdot\bar\varphi_0}\vert 0,0;\alpha_0\rangle,
\label{eqn:ketground}
\eea
where
\beq
V_\alpha(z) = e^{i\alpha\cdot\phi(z)}, \;\;
\bar V_{\bar\alpha}(\bar z)=e^{i\bar\alpha\cdot\bar\phi(\bar z)}.
\eeq
The Fock spaces are generated on these ground states by operating with the negative mode operators $a^{(i)}_{n<0}$ and $\bar a^{(i)}_{n<0}$.
The corresponding bra ground states are given by
\beq
\langle\alpha,\bar\alpha;\alpha_0\vert
= \langle 0,0;\alpha_0\vert e^{-i\alpha\cdot\varphi_0}e^{-i\bar\alpha\cdot\bar\varphi_0}.
\label{eqn:braground}
\eeq
Note the difference of signs in the exponents of (\ref{eqn:ketground}) and (\ref{eqn:braground}).
The coefficients of $i\varphi_0$ ($i\bar\varphi_0$) in the exponents are the left-moving (right-moving) zero-mode momenta [holomorphic (anti-holomorphic) Coulomb-gas charges in the Coulomb-gas language] and are subject to momentum conservation [charge neutrality condition in Coulomb-gas].
The ket states $\vert\alpha,\bar\alpha;\alpha_0\rangle$ are thus interpreted to have momenta $\alpha$ and $\bar\alpha$, whereas the momenta of the bra states $\langle\alpha,\bar\alpha;\alpha_0\vert$ are
$-\alpha$ and $-\bar\alpha$.
With these definitions the momentum conservation demands the ground states being orthogonal, and we normalise them as 
$\langle\alpha,\bar\alpha;\alpha_0\vert\beta,\bar\beta;\alpha_0\rangle
=\delta_{\alpha\beta}\delta_{\bar\alpha\bar\beta}$.
The value of $\alpha_0$ is common since it is fixed by (\ref{eqn:alpha0andk}).

Expanding the stress tensor (\ref{eqn:stresstensor}) in powers of $z$, we find the Virasoro operators
\beq
L_n=\frac 12\sum_{m\in{\mathbb Z}}:a_m\cdot a_{n-m}:-2\alpha_0(n+1)\rho\cdot a_n.
\eeq
We look for states\footnotemark[3] $\vert B\rangle$
\footnotetext[3]{The $B$ stands for {\em b}oundary here.
We hope no confusion arises with the B-type boundary states.}
satisfying the Ishibashi condition
\beq
(L_n-\bar L_{-n})\vert B\rangle=0,
\label{eqn:ishibashicond}
\eeq
in the Fock spaces introduced above.
We make a coherent state ansatz \cite{Caldeira:2003zz},
\bea
\vert B(\alpha,\bar\alpha,\Lambda)\rangle
&=&C_\Lambda
\vert\alpha,\bar\alpha;\alpha_0\rangle,\\
C_\Lambda&=&\prod_{m>0}\exp\left(\frac 1m a_{-m}\cdot \Lambda\cdot\bar a_{-m}\right),
\eea
where $\Lambda$ acts on the three components of the bosonic field as a
$3\times 3$ matrix.
Explicit calculation using the
Baker-Campbell-Hausdorff formula shows that
the boundary condition (\ref{eqn:ishibashicond}) implies
\bea
&& \Lambda^T \cdot \Lambda=I,\label{eqn:cond1}\\
&& \Lambda\cdot\rho+\rho=0,\label{eqn:cond2}\\
&& \Lambda^T\cdot\alpha+4\alpha_0\rho-\bar\alpha=0.
\label{eqn:cond3}
\eea
From (\ref{eqn:cond1}) and (\ref{eqn:cond2}) we find that $\Lambda$ is of the form
\beq
\Lambda=
-\left(
\begin{array}{cc}
1&\begin{array}{cc}0&0\end{array}\\\begin{array}{c}0\\0\end{array}&M
\end{array}
\right),
\label{eqn:lambda}
\eeq
where $M$ is a $2\times 2$ orthogonal matrix.
As the right-moving momenta are related to the left-moving ones
through (\ref{eqn:cond3}), we introduce an abbreviated notation of states
\beq
\vert B(\alpha, \Lambda) \rangle
\equiv\vert B(\alpha,\Lambda^T\cdot\alpha+4\alpha_0 \rho, \Lambda)\rangle,
\label{eqn:bket}
\eeq
where $\Lambda$ is given by (\ref{eqn:lambda}).
The corresponding bra states are
\beq
\langle B(\alpha, \Lambda) \vert
=\langle\alpha,\Lambda^T\cdot\alpha+4\alpha_0 \rho; \alpha_0\vert C_\Lambda^T,
\label{eqn:bbra}
\eeq
where
\beq
C_\Lambda^T=\prod_{m>0}\exp\left(\frac 1m a_{m}\cdot \Lambda\cdot\bar a_{m}\right).
\eeq
The overlaps between these states are computed using the free-field representation as
\bea
&&\langle B(\alpha, \Lambda) \vert q^{\frac 12 (L_0+\bar L_0-\frac{c}{12})}
\vert B(\beta, \Lambda')\rangle\nn\\
&&=\frac{q^{\frac 12 (\alpha-2\alpha_0\rho)^2-\frac 18}}{\prod_{m>0}
\det(1-q^m\Lambda^T\cdot\Lambda')}
\delta_{\alpha,\beta}\delta_{\Lambda^T\cdot\alpha,\Lambda'^T\cdot\beta}.
\label{eqn:boverlaps}
\eea

The states (\ref{eqn:bket}), (\ref{eqn:bbra}) constructed above certainly
satisfy the Ishibashi condition (\ref{eqn:ishibashicond}).
However, they are not boundary states of the CFT. Having no restriction
on $\alpha$ means that some of the states are spurious. Physical
states must be part of the cohomology of the screening operators,
{\em i.e.} elements of the Felder complex. This is related to the
requirement that boundary states must give overlaps
which are partition functions for (irreducible)
representations of the chiral algebra.
Thus, we will instead consider the states $\vert \Lambda; j\rishi$
for $\Lambda$ of (\ref{eqn:lambda}) and each isospin $j$ of $SU(2)$,
obtained by summing the charges $\alpha$ over the Felder complex.
Explicitly,
\bea
\vert \Lambda; j \rishi = \sum_{\alpha\in\Gamma_j}\kappa_{n_1}
\vert B(\alpha, \Lambda)\rangle,
\label{eqn:su2ishibashi}
\eea
where $\kappa_{n_1}$ is a phase factor related to its bra counterpart $\kappa'_{n_1}$ through (\ref{eqn:kapparel}) below\footnotemark[4], and the lattice $\Gamma_j$ is
\footnotetext[4]{
These phase factors may be determined in principle by BRST invariance of the boundary states \cite{Parkhomenko:2001ki}. 
Our computation of correlation functions on disk topology below does not depend on such details since this factor can be absorbed in the vacuum normalisation on the disk.  
}
\bea
\Gamma_j &:& \alpha^{(1)}=-j\sqrt{\frac{2}{k+2}}-n_1\sqrt{2(k+2)},\nn\\
&&\alpha^{(2)}=-im\sqrt{\frac 2k}-in_2\sqrt{2k},\nn\\
&&\alpha^{(3)}=m\sqrt{\frac 2k}+n_3\sqrt{2k},
\label{eqn:gammaj}
\eea
where
\bea
&&n_1, n_2 \in \frac 12 {\mathbb Z}, \;\; n_3, \in {\mathbb Z}, \nn\\
&&n_1-n_2 \in {\mathbb Z}, \nn\\
&&n_1\ge |n_2|, \;\; -n_1>|n_2|, \nn\\
&&m\in \frac 12 {\mathbb Z}, \;\; 0\leq m \leq \frac{2k-1}{2}, \;\; j-m\in {\mathbb Z}.
\label{eqn:latticecond}
\eea
The lattice points of (\ref{eqn:gammaj}) contain a part indexed by
$(j,m)$, equal to the labels (\ref{eqn:alphajm})
of the $SU(2)$ primaries, and a part indexed by $(n_1,n_2,n_3)$,
corresponding to periodic identifications. In fact, the
$\alpha^{(3)}$ sector is precisely the $U(1)_k$ periodic lattice
of (1). 
The lattice sums for $n_1$ and $n_2$ are those appeared in the parafermion character formula (\ref{eqn:pfchar}). The $\alpha^{(1)},\alpha^{(2)}$ sector is thus identified as the parafermion part.

The bra states $\lishi \Lambda; j\vert$ are constructed similarly but with a phase factor $\kappa'_{n_1}$ in place of $\kappa_{n_1}$, satisfying
\beq
\kappa'_{n_1}\kappa_{n_1}=(-1)^{2n_1}{\rm sign}(n_1),
\label{eqn:kapparel}
\eeq
where we define ${\rm sign}(0)=1$ as before.
With these definitions it is obvious from (\ref{eqn:boverlaps}) and the character formulae (\ref{eqn:pfchar}), (\ref{eqn:su2char}) that the overlaps between these states give the $SU(2)_k$ characters,
\beq
\lishi \Lambda; j\vert q^{\frac 12 (L_0+\bar L_0-\frac{c}{12})}\vert \Lambda; j' \rishi
=\chi_j^{SU(2)}(\tau)\delta_{jj'}.
\eeq
The states $\vert\Lambda; j\rishi$ are hence regarded as the Ishibashi states of the $SU(2)_k$ model
\footnotemark[5].
\footnotetext[5]{There remains some ambiguity due to the reflection symmetry of the Weyl group, which can be removed as \cite{Ishikawa:2000ez, Kawai:2002pz} or \cite{Caldeira:2004jy}.}

These states are indexed by $\Lambda$ as well as $j$.
$\Lambda$ is a generalisation of the sign difference in (\ref{eqn:U1AIshi}) and (\ref{eqn:U1BIshi})
which distinguishes the A- and B-type $U(1)_k$ Ishibashi states.
So we expect that $\Lambda$ specifies the type of the boundary states in the $SU(2)_k$ theory.
Indeed, with an explicit calculation using the bosonic
expressions of the $SU(2)$ currents (\ref{eqn:J}) and (\ref{eqn:Jbar}),
we can show that $\Lambda$ is related to the current gluing conditions.
In the present coordinates $z,\bar z$ the boundary is the unit
circle $z=\bar z^{-1}$. Identifying
\beq
M=\tilde M,
\eeq
we see that on the boundary the Ishibashi states satisfy
%
%
\bea
&&\left[zJ^\pm(z)-\bar z\Omega\bar J^\mp(\bar z)\right]
\vert \Lambda; j\rishi = 0,
\label{eqn:glue1}\\
&&\left[zJ^3(z)-\bar z\Omega\bar J^3(\bar z)\right]
\vert \Lambda; j\rishi = 0.
\label{eqn:glue2}
\eea

If we map the unit disk onto the upper half plane, the
above conditions reduce to the standard form
of the gluing automorphisms in the open string picture,
\beq
J^\pm+\Omega \bar J^\mp=0, \;\; J^3+\Omega \bar J^3=0.
\label{eqn:Jgluing}
\eeq
Note that these gluing conditions reduce to
\beq
J^\pm+\bar J^\mp=0 \ , \  J^3+\bar J^3=0
\eeq
when $M={\rm diag}(1, 1)$ and
\beq
J^\pm+\bar J^\pm=0 \ , \ J^3-\bar J^3=0
\eeq
when $M={\rm diag}(-1, -1)$, but cannot be written in
simple forms in other cases.

In order to look into this in more detail, let us first
focus on the A-type states of the $SU(2)_k$ model, which are
characterised by the trivial gluing conditions of the currents $J^a$.
Consider the half-plane geometry where the A-type boundary condition is imposed
on the real axis and use the mirroring of \cite{Cardy:1984bb}.
The trivial current gluing conditions imply that the antiholomorphic currents $\bar J^a$ (on the "lower half plane") are analytical continuations of the holomorphic currents $J^a$ on the upper half plane.
This allows us to map the CFT on the half plane to a chiral CFT on the full
plane. A $p$-point correlation function on the half plane is then
equivalent to a $2p$-point function on the full plane.
In particular, consider a one point function of a primary field $\Phi_{j,m}(z,\bar z)$ on the upper half plane, which is written as a two point function on the full plane:
\beq
\langle\Phi_{j,m}(w,\bar w)\rangle_{{\rm UHP},A}
= \langle \Phi_{j,m}(w)\Phi_{j,-m}(w^*)\rangle_{\rm FP}.
\label{eqn:uhp1pf}
\eeq
Let us evaluate the left hand side in the free field representation.
We first map the upper half plane to the unit disk. Then we
represent the primary operator $\Phi_{j,m}(z,\bar z)$ by $V_{j,m}(z) \bar V^\dag_{j,m}(\bar z)$
using the free field representations (\ref{eqn:su2vertex}), and represent
the boundary with the boundary state $\lishi \Lambda ; j\vert$. The lhs
is thus proportional to
\beq
\lishi\Lambda; j\vert V_{j,m}(z)\bar V^\dag_{j,m}(\bar z)
\vert 0,0;\alpha_0\rangle \ .
\label{eqn:disk1pf}
\eeq
Note that we could as well have used one of the other
vertex operator representations of (\ref{eqn:su2vertex}).
The other representations would also lead to the same result as below, but
require insertion of screening operators, making the calculation a bit longer.
The disk one point function (\ref{eqn:disk1pf}) vanishes unless the left- and right-moving momenta are separately conserved (modulo screening charges),
\bea
&&-\alpha+\alpha_{j,m}=0,\\
&&-\Lambda^T\cdot\alpha-4\alpha_0\rho+\alpha^\dag_{j,m}=0,
\eea
where $\alpha$ is one of the boundary momenta (Coulomb-gas charges) summed over in (\ref{eqn:su2ishibashi}).
It is obvious that the above conditions are satisfied for $\alpha=\alpha_{j,m}$ only when
$\Lambda={\rm diag}(-1,1,1)$.
As the right hand side of (\ref{eqn:uhp1pf}) is clearly non-vanishing, we conclude that A-type boundary states correspond to
$\Lambda={\rm diag}(-1,1,1)$,
and hence the A-type Ishibashi states in the $SU(2)_k$ theory are written as
\beq
\vert A; j\rishi= \vert \Lambda={\rm diag}(-1,1,1); j\rishi
\label{eqn:su2AIshi}
\eeq
(In fact, there is some freedom left, which we explain in the next
subsection).
With these Ishibashi states the A-type Cardy states are constructed in the usual way as
\beq
\vert \widetilde{A; j}\rangle =  \sum_{j'}\frac{S_{jj'}}{\sqrt{S_{0j'}}}\vert A; j'\rishi,
\label{eqn:su2ACardy}
\eeq
where $S_{\lambda \mu}$, $\lambda, \mu = 0, \frac 12, \cdots \frac k2$, 
is the $SU(2)_k$ modular S-matrix,
\beq
S_{\lambda \mu}=\sqrt{\frac{2}{k+2}}\sin\left(\frac{\pi (2\lambda+1)(2\mu+1)}{k+2}\right).
\eeq


\subsection{One point functions and rotation of branes}


In the previous subsection we identified the A-branes by
the condition (\ref{eqn:uhp1pf}), namely the universal coupling between
the Ishibashi states and the bulk primary operators.
The obtained states agree with what is known in the literature.
In particular, one may reproduce overlaps between these states by explicit
free field calculations.
The above representation for the A-branes (\ref{eqn:su2AIshi}) is
however not general enough since the branes may be rotated within the group
without changing their properties.
We argue that such rotations are implemented in the free field
representation by rotations between $\bar\phi^{(2)}$ and $\bar\phi^{(3)}$, 
namely the transformation $\omega$ of (\ref{eqn:omega}) which keeps
the value of the $\det \tilde M$ constant.

In deriving (\ref{eqn:su2AIshi}) we have used the convention that the non-chiral primary operators have equal (up to the conjugation $\alpha_{j,m}\leftrightarrow\alpha^\dag_{j,m}$ ) left- and right-momenta.
This condition can be relaxed, because instead of $\bar V_\alpha(\bar z)$ we may use rotated antiholomorphic vertex operators,
\beq
\Omega\bar V_\alpha(\bar z)\equiv\exp\left( i\alpha\cdot\omega\bar\phi(\bar z)\right).
\label{eqn:RotatedVBar}
\eeq
Now, upon identifying $\tilde M$ of $\omega$ and $M$ of $\Lambda$, we can see that charge neutrality is satisfied in the one point functions
\beq
\lishi\Lambda; j\vert V_{j,m}(z)\Omega\bar V_{j,m}^\dag(\bar z)\vert 0,0; \alpha_0\rangle
\eeq
for {\em all} $\Lambda=-(1,M)$.
This indicates that bulk primary fields transformed by $(1\otimes\omega)$ always feel the states $\lishi\Lambda; j\vert$ as A-branes.
We may thus allow a rotation of the right-moving part of the bulk primary field and regard the branes
rotated from (\ref{eqn:su2AIshi}), that is, $\vert\Lambda; j\rishi$ with $\det M=+1$, as A-branes.
In particular, $\Lambda={\rm diag}(-1,-1,-1)$ also represents A-branes
(with appropriately rotated bulk operators).
We will use this in the next subsection.


\subsection{Parafermion and B-type $SU(2)_k$ boundary states}


In the $SU(2)_k$ WZNW model there are known to be other kinds of branes called B-type, 
which are constructed from A-branes by orbifolding and T-duality \cite{Maldacena:2001ky}.
Before discussing the B-branes in $SU(2)_k$, let us consider the A- and B-type boundary states of
the parafermion model $SU(2)_k/U(1)_k$.

In the general $SU(2)_k$ Ishibashi state formula
(\ref{eqn:su2ishibashi}), one could clearly identify a sector
corresponding to the parafermions, and a sector corresponding to
the $U(1)_k$. Stripping off the latter contribution then gives
the Fock space representation of the A-type
parafermion Ishibashi states,
\bea
&&\vert A; j,n\rishi_{PF}\nn\\
&&=\prod_{m>0}\exp\left\{-\frac 1m a_{-m}^{(1)}\bar a_{-m}^{(1)}
+\frac 1m a_{-m}^{(2)}\bar a_{-m}^{(2)}\right\}\nn\\
&&\times\!\!\!\!\!\!\!\!\!\!\!\!\sum_{\alpha^{(1)}, \alpha^{(2)} \in \Gamma^{PF}_{j,n}}\!\!\!\!\!\!\!\!\!\!\!
\kappa_{n_1}
\!\vert \alpha^{(1)}\!\!, \alpha^{(2)}\!\!, \bar\alpha^{(1)}\!=\!4\alpha_0\rho-\alpha^{(1)}\!\!, \bar\alpha^{(2)}\!=\!\alpha^{(2)}; \alpha_0\rangle,\nn\\
\label{pfAishi}
\eea
where $\kappa_{n_1}$ is the same phase factor as in the $SU(2)_k$ case.
The parafermion charge summation goes over the sublattice of (\ref{eqn:gammaj}),
\bea
\Gamma^{PF}_{j,n}&:&
\alpha^{(1)}=-j\sqrt{\frac{2}{k+2}}-n_1\sqrt{2(k+2)},\nn\\
&&\alpha^{(2)}=-in\frac{1}{\sqrt{2k}}-in_2\sqrt{2k},
\label{eqn:pfgamma1}
\eea
where
\bea
&&n_1, n_2 \in \frac 12 {\mathbb Z}, \nn\\
&&n_1-n_2 \in {\mathbb Z}, \nn\\
&&n_1\ge |n_2|, \;\; -n_1>|n_2|.
\label{eqn:pfgamma2}
\eea

Consider then the expression in \cite{Maldacena:2001ky}
for the A-type $SU(2)_k$ Ishibashi states as a combination of the
A-type parafermionic and $U(1)_k$
states,
\beq
\vert A;j\rishi = \sum_{n=0}^{2k-1}\frac{1+(-1)^{2j+n}}{2}
\vert A; j,n\rishi_{PF} \vert A; n\rishi_{U(1)}.
\eeq
If we substitute into this the expressions (\ref{pfAishi}) and
(\ref{eqn:U1AIshi}), we recover the formula
(\ref{eqn:su2ishibashi}). The projection $\frac{1+(-1)^{2j+n}}{2}$ enforces
the constraint $j-m\in {\mathbb Z}$ of (\ref{eqn:latticecond}).

The B-type Ishibashi states in the parafermion theory are
defined in \cite{Maldacena:2001ky} using a
operator $e^{i\pi \bar J^1_0}$ which changes the sign of the $\bar J^3_0$
eigenvalue, as
\bea
&&(1\otimes e^{i\pi\bar J^1_0})\vert A;j\rishi\nn\\
&&=\sum_{n=0}^{2k-1}
\frac{1+(-1)^{2j+n}}{2}\vert B; j, n\rishi_{PF}
\vert B; n\rishi_{U(1)}.\nn\\
\label{eqn:dcmp2}
\eea
In the free-field language, the action of the operator
$(1+e^{i\pi\bar J^1_0})$ (which flips the sign
of $n$ in the above) is equivalent to replacing $M$ by $-M$.
Hence using the free field representation (\ref{eqn:su2AIshi}) the left hand
side of (\ref{eqn:dcmp2}) is the $SU(2)_k$ A-type Ishibashi states
with $\Lambda={\rm diag}(-1,-1,-1)$.
Since the $U(1)_k$ part of these states is clearly the B-type, we can mod it out according to (\ref{eqn:dcmp2}) and find,
\bea
&&\vert B; j,n\rishi_{PF}\nn\\
&&=\prod_{m>0}\exp\left\{-\frac 1m a_{-m}^{(1)}\bar a_{-m}^{(1)}
-\frac 1m a_{-m}^{(2)}\bar a_{-m}^{(2)}\right\}\nn\\
&&\times\!\!\!\!\!\!\!\!\!\!\!\sum_{\alpha^{(1)}, \alpha^{(2)} \in \Gamma^{PF}_{j,n}}
\!\!\!\!\!\!\!\!\!\!\!\!\!\kappa_{n_1}\vert \alpha^{(1)}\!, \alpha^{(2)}\!, \bar\alpha^{(1)}\!=\!4\alpha_0\rho-\alpha^{(1)}\!, \bar\alpha^{(2)}\!=\!-\alpha^{(2)}; \alpha_0\rangle, \nn\\
\eea
where the lattice summation is the same as for
the A-type states (\ref{eqn:pfgamma1}), (\ref{eqn:pfgamma2}).
The Cardy states of the parafermion theory are constructed in the standard
manner, using the above free field representation for the Ishibashi states.

Finally, the B-branes of the $SU(2)_k$ theory  are constructed from the A-type
parafermion states and B-type $U(1)_k$ states, or from the B-type parafermion
states and A-type $U(1)_k$ states \cite{Maldacena:2001ky}.
For example, one may construct the B-type
Cardy states of the $SU(2)_k$ model as
\bea
&&\vert \widetilde{B; j,\eta}\rangle\nn\\
&&=\sqrt k \sum_{j'=0}^{k/2}\frac{S_{jj'}}{\sqrt{S_{0j'}}}\left[
\frac{1+(-1)^{2j'}}{2}\vert A; j', 0\rishi_{PF}\vert B; 0\rishi_{U(1)}\right.\nn\\
&&\left.+(-1)^{2j}\eta\frac{1+(-1)^{2j'+k}}{2}\vert A; j', k\rishi_{PF}\vert B; k\rishi_{U(1)}\right],
\label{eqn:su2BCardy}
\eea
where $\eta=\pm 1$.
In the above case the states have $\Lambda={\rm diag}(-1, 1, -1)$ but in a similar construction from B-type parafermion states and A-type $U(1)_k$ states we have $\Lambda={\rm diag}(-1,-1,1)$.
These two cases are connected by a rotation within the group, and hence B-type $SU(2)_k$ boundary
states are characterized by $\det M=-1$.
In summary, boundary states with $\det M = +1$ are considered as A-branes and $\det M=-1$ as B-branes in the $SU(2)_k$ model.

We comment on a subtlety arising in the distinction of A- and B-type branes.
The A- and B-type boundary states constructed above have desired properties that A-type branes couple to bulk operators (with any magnetic quantum number $m$) universally, whereas B-type branes couple only to operators with $m=0$ up to reflection symmetry.
These are fundamental properties which distinguish A- and B-branes.
However, we might as well use a (slightly unnatural) convention such that the bulk primary fields are constructed from a holomorphic part and its "twisted" antiholomorphic counterpart, that is, (\ref{eqn:RotatedVBar}) with $\tilde M$ connected to ${\rm diag}(+1,-1)$ by rotation.
In this case the roles of branes with $\det M=\pm 1$ are completely exchanged, and we must regard $\det M=+1$ states as type B and $\det M=-1$ as type A since the A- and B-branes are entirely symmetric (e.g. the overlaps between them) except for the coupling with bulk operators.
In this sense the definitions of A- and B-branes are relative, and depend on how to define bulk non-chiral operators.


\section{Correlation functions, chiral blocks and boundary states}
The Coulomb-gas computation of WZNW correlation functions was developed in 
\cite{Dotsenko:1990ui, Dotsenko:1991zb, Gerasimov:1990fi, Nemeschansky:1989wg, Bilal:1989py} 
and was extended to higher topologies in 
\cite{Bernard:1990iy, Jayaraman:1990tu, Frau:1990pt, Nemeschansky:1991rx, Konno:1992vm}.
In this section we apply this technique to the computation on the disk topology, where the boundary condition is represented by the boundary states which have been discussed  in the previous section.
As physical boundary conditions are realized by the Cardy states, a p-point function on the unit disk which we shall consider is given by
\bea
&&\langle{\mbox{Cardy state}}\vert V_1(z_1)\bar V_1(\bar z_1) \cdots V_p(z_p)\bar V_p(\bar z_p)\nn\\
&&\;\;\;\times(\mbox{screening operators}) \vert 0,0;\alpha_0\rangle,
\label{eqn:diskcorrel1}
\eea
where $\vert 0,0;\alpha_0\rangle$ is the $SL(2,{\mathbb C})$ invariant vacuum at the centre of the disk.
The free field representations of the boundary states, the primary operators and the 
screening operators have already been given.
Thus the above expression can be straightforwardly evaluated in the free field formalism, i.e. with repeated use of the Heisenberg algebra and the Baker-Campbell-Hausdorff formula.

A few technical remarks for the actual computation are in order\cite{Kawai:2002pz, Caldeira:2003zz}.
Firstly, we are allowed to insert different numbers of screening operators in holomorphic and anti-holomorphic sectors, although this might seem odd from the mirroring (or doubling) picture of \cite{Cardy:1984bb}.
The reason for this is that the mirroring argument is based on the analytic continuation of the holomorphic and antiholomorphic currents but the screening operators have by construction trivial effects on the currents of the chiral algebra; in other words, the mirror does not see the screening operators.
The second point is that, as the boundary has charges, the neutrality condition of the Coulomb-gas charges must now take the contributions from the boundary into account.
The Cardy states constructed in the last section are linear sums of the states
$\langle B(\alpha,\Lambda)\vert$ which have a holomorphic charge $-\alpha$ and an anti-holomorphic charge $-(\Lambda^T\cdot\alpha+4\alpha_0\rho)$ (recall that $\langle\alpha,\bar\alpha;\alpha_0\vert$ and $\vert\alpha,\bar\alpha;\alpha_0\rangle$ have opposite charges).
The correlation function (\ref{eqn:diskcorrel1}) thus reduces to a sum of the amplitudes,
\bea
&&\langle B(\alpha,\Lambda)\vert V_1(z_1)\bar V_1(\bar z_1)\cdots V_p(z_p)\bar V_p(\bar z_p)\nn\\
&&\;\;\;\times(\mbox{screening operators})\vert 0,0;\alpha_0\rangle,
\label{eqn:diskcorrel2}
\eea
whose holomorphic and anti-holomorphic charges must be independently neutral:
\bea
&&-\alpha+\alpha_1+\cdots+\alpha_p\nn\\
&&\;\;+(\mbox{holomorphic screening charges})=0,\\
&&-(\Lambda^T\cdot\alpha+4\alpha_0\rho)+\bar\alpha_1+\cdots+\bar\alpha_p\nn\\
&&\;\;+(\mbox{antiholomorphic screening charges})=0.
\eea
Otherwise the amplitudes vanish.
The neutrality of charges corresponds to consistency of fusion rules among the primary operators\cite{Felder:1989zp}, and non-vanishing amplitudes (\ref{eqn:diskcorrel2}) are interpreted as chiral blocks\cite{Kawai:2002pz}.

Below we shall give sample calculations based on these observations.
We focus on simple two point functions of the $SU(2)_k$ model and explain the method for both 
A- and B-type Cardy states.


\subsection{A-branes in $SU(2)_k$ model}


Let us first consider a general two point function of bulk primary operators
$\Phi_{j_1,m_1}(z_1,\bar z_1)$ and $\Phi_{j_2,m_2}(z_2,\bar z_2)$ on the unit disk.
On the boundary of the disk we assume the A-type Cardy states $\langle\widetilde{A; j}\vert$.
The calculation for the A-type boundary is entirely straightforward and parallel to the corresponding four point case on the full plane (e.g. \cite{Gerasimov:1990fi}).
This of course is expected since the trivial current gluing conditions implied by the A-type boundary allow analytic continuation of the chiral currents to the full plane and then the correlation function we are considering must satisfy the same Knizhnik-Zamolodchikov equations as for the chiral four point function on the full plane.
We shall choose the vertex operator representation of the bulk operators as
\bea
\Phi_{j_1,m_1}(z_1,\bar z_1)&:& V_{j_1,m_1}(z_1)\bar V_{j_1,m_1}(\bar z_1),\nn\\
\Phi_{j_2,m_2}(z_2,\bar z_2)&:& V_{j_2,m_2}(z_2)\bar V^\dag_{j_2,m_2}(\bar z_2),
\label{eqn:2ptvertex}
\eea
see section 2 for their definitions.
We might choose other representations from the ones listed in (\ref{eqn:su2vertex}) but they are equivalent (and involve unnecessarily many screening operators in following computations).
Given holomorphic and antiholomorphic boundary charges $-\alpha$ and 
$-\bar\alpha=-4\alpha_0\rho-\Lambda^T\cdot\alpha$ 
(note that we have $\Lambda={\rm diag}(-1,+1,+1)$ for the A-branes now), 
the sums of the bulk and boundary Coulomb-gas charges in (\ref{eqn:diskcorrel2}) are,
\bea
{\rm Holomorphic}&:&-\alpha+\alpha_{j_1,m_1}+\alpha_{j_2,m_2},\\
{\rm Antiholomorphic}&:&-4\alpha_0\rho-\Lambda^T\cdot\alpha+\alpha_{j_1,m_1}+\alpha^\dag_{j_2,m_2}.\nn\\
\eea
Now we ask whether these charges are screenable, that is, whether it is possible to neutralize them by inserting screening operators.
It is a simple exercise to check that the both sectors are neutralized by $n$ holomorphic screening operators
\beq
Q_1=\oint dz \del\phi^{(2)} e^{i\sqrt{\frac{2}{k+2}}\phi^{(1)}},
\eeq
and $(l-n)$ anti-holomorphic screening operators,
\beq
\bar Q_1=\oint d\bar z \bar\del\bar\phi^{(2)} e^{i\sqrt{\frac{2}{k+2}}\bar\phi^{(1)}},
\eeq
where $l=2j_1$ and $n=0, 1, \cdots, l$.
For each $n$, the charge $\alpha$ on the boundary is found to be
\bea
\alpha&=&\left(
(n-j_1-j_2)\sqrt{\frac{2}{k+2}}, \right.\nn\\
&&\left. -i(m_1+m_2)\sqrt{\frac 2k}, (m_1+m_2)\sqrt{\frac 2k}
\right).
\label{eqn:boundaryalpha}
\eea
Note that the values of $\alpha$ reflect the fusion of the $SU(2)_k$ primary fields,
\beq
\Phi_{j_1,m_1}\times\Phi_{j_2,m_2}
=\sum_{\stackrel{\scriptstyle j=|j_1-j_2|}{j_1+j_2+j\in{\mathbb Z}}}
^{\min (j_1+j_2, k-j_1-j_2)}\Phi_{j,m=m_1+m_2}.
\eeq
The neutrality of charges implies that the resulting intermediate state $\Phi_{j,m}$ couples to the boundary, only via the corresponding Ishibashi state $\lishi A; j\vert$ (figure 2).

\begin{figure}
\setlength{\unitlength}{3pt}

~~~\begin{picture}(20,30)
{\thicklines
\put(0,15){\line(1,0){20}}}
{\thinlines
\put(10,20){\line(0,-1){10}}
\put(10,20){\line(1,1){5}}
\put(10,20){\line(-1,1){5}}
\put(10,10){\line(1,-1){5}}
\put(10,10){\line(-1,-1){5}}}
\put(5,25){\circle*{2}}
\put(15,25){\circle*{2}}
\put(5,5){\circle*{2}}
\put(15,5){\circle*{2}}
\put(1,28){$\Phi_{j_2,m_2}$}
\put(15,28){$\Phi_{j_1,m_1}$}
\put(1,1){$\bar\Phi_{j_2,m_2}$}
\put(16,1){$\bar\Phi_{j_1,m_1}$}
\put(11,17){$\Phi_{j,m}$}
\end{picture}

\caption{Chiral block for a two point function on the half plane. 
The intermediate state $\Phi_{j,m}$ occurring as a fusion product of $\Phi_{j_1,m_1}$ 
and $\Phi_{j_2,m_2}$ couples to the corresponding Ishibashi state on the boundary.}
\end{figure}
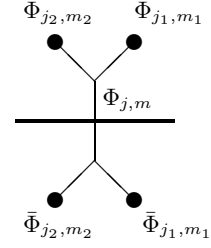
Using the expression of the A-type Cardy states (\ref{eqn:su2ACardy}) the two point function is written as a linear combination of chiral blocks,
\bea
&&\langle\Phi_{j_1,m_1}(z_1,\bar z_2)\Phi_{j_2,m_2}(z_2,\bar z_2)\rangle_{{\rm disk}, \widetilde{A,j}}\nn\\
&&=\frac{\sqrt{S_{00}}}{S_{j0}}\sum_{j'}\frac{S_{jj'}}{\sqrt{S_{0j'}}}I^A_{j',m},
\eea
where $m=m_1+m_2$ and 
\bea
&&I^A_{j',m}=\langle B(\alpha,\Lambda)\vert
V_{j_1,m_1}(z_1)\bar V_{j_1,m_1}(\bar z_1) \nn\\
&&\;\;\;\times V_{j_2,m_2}(z_2)\bar V^\dag_{j_2,m_2}(\bar z_2)
Q_1^n \bar Q_1^{l-n}\vert 0,0;\alpha_0\rangle.
\label{eqn:chiralblock}
\eea
The overall factor
$\sqrt{S_{00}}/S_{j0}=\langle\widetilde{A; j}\vert 0,0;\alpha_0\rangle^{-1}$ comes from the normalization of the vacuum.
The boundary charge $\alpha$ is as in (\ref{eqn:boundaryalpha}) and $n=j_1+j_2-j'$, $\Lambda={\rm diag}(-1,+1,+1)$.
The screening charges yield $n$-tuple integration in the holomorphic 
and $(l-n)$-tuple integration in the anti-holomorphic sector.
The contours are initially those of Felder's, but following \cite{Felder:1989zp} one may deform them
into
\beq
\int_{C_1}dt_1\cdots \int_{C_n}dt_n \int_{S_1}ds_1\cdots \int_{S_{l-n}}ds_{l-n},
\eeq
where $C_i$ and $S_i$ are as shown in figure 3.
The chiral blocks (\ref{eqn:chiralblock}) are then written by the standard Dotsenko-Fateev integrals.

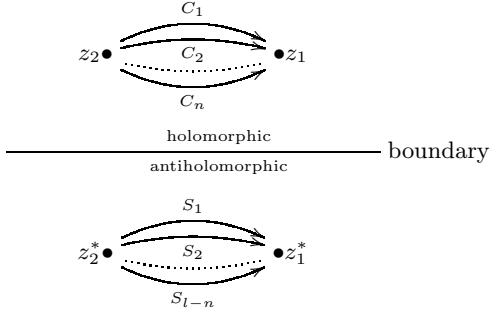
\begin{figure}
\hspace{10mm}
\xymatrix{
{}&{z_2\bullet}\ar@{->}@/^1pc/[rr] ^{C_1} \ar@/^/[rr]_{C_2} \ar@{.}@/_/[rr] \ar@/_1pc/[rr]_{C_n}&{}&\bullet z_1\\
{}\ar@{-}[rrrr]^{\rm holomorphic}_{\rm antiholomorphic}&{}&{}&{}&{\rm boundary}\\
{}&{z_2^*\bullet}\ar@/^1pc/[rr]^{S_1} \ar@/^/[rr]_{S_2} \ar@{.}@/_/[rr] \ar@/_1pc/[rr]_{S_{l-n}}&{}&\bullet z_1^*\\
}
\caption{Integration contours of chiral blocks for A-branes. }
\end{figure}

These involve $l$ screening operators in total and hence there are $l+1$ independent choices of contours, depending on how many of them are in the holomorphic sector.
As the configuration of the screening charges is now related to the Ishibashi states through (\ref{eqn:boundaryalpha}), we expect that for each Ishibashi state there corresponds one of the $l+1$ independent solutions\cite{Christe:1987cy, Gerasimov:1990fi} of the Knizhnik-Zamolodchikov equations.
Let us see this in a specific example where $j_1=j_2=1/2$, $m_1=1/2$ and $m_2=-1/2$.
In this case the number of the screening charges is $l=1$, and hence $n$ can be either $1$ or $0$.
When $n=1$, from (\ref{eqn:boundaryalpha}) we see that $\alpha=(0,0,0)$, which is in the lattice sum of the A-type Ishibashi state $\lishi A;j=0\vert$.
Using the free field formalism we may express the chiral block as
\bea
I_{0,0}^A&=&{\rm const.}\times z_1^{\frac{1}{2(k+2)}}\bar z_1^{-\frac{3}{2(k+2)}}
(1-\xi)^{\frac{1}{2(k+2)}} \nn\\
&&\oint dt (z_1-t)^{-\frac{1}{k+2}}t^{-\frac{1}{k+2}}(1-\bar z_1 t)^{-\frac{1}{k+2}}\nn\\
&&\times\left\{\frac 1t +\frac{1}{z_1-t}-\frac{\bar z_1}{1-\bar z_1 t}\right\},
\eea
where $\xi \equiv z_1\bar z_1$ is a cross ratio and
we have used the global conformal invariance to set $z_2=\bar z_2=0$.
Changing the holomorphic contour into
\beq
\oint dt \rightarrow \int_0^{z_1} dt,
\eeq
the chiral block is found in the form using a hypergeometric function,
\beq
I_{0,0}^A=C_{0,0}\xi^{-\frac{3}{2(k+2)}}(1-\xi)^{\frac{1}{2(k+2)}}
F(\frac{1}{k+2},\frac{k+1}{k+2},\frac{k}{k+2};\xi).
\label{eqn:A00block}
\eeq
The normalization constant $C_{0,0}$ is identified as the 3-point coupling constant in the operator product expansion,
\bea
&&\Phi_{1/2,1/2}(z,\bar z)\Phi_{1/2,-1/2}(0,0)\nn\\
&&\;=C_{0,0}\Phi_{0,0}(0,0)\vert z\vert^{-\frac{3}{k+2}}+C_{1,0}\Phi_{1,0}(0,0)
\vert z\vert^{\frac{1}{k+2}}\cdots,\nn\\
\eea
In the off-boundary limit the chiral block agrees with the bulk two point function.
For $n=0$ we have one screening charge in the antiholomorphic sector and (\ref{eqn:boundaryalpha}) becomes
$\alpha=(-\sqrt{\frac{2}{k+2}}, 0,0)$.
This boundary charge is included in the Ishibashi state $\lishi A; j=1\vert$, indicating the intermediate state of the chiral block being $\Phi_{1,0}$.
The evaluation of the chiral block goes similarly as above, giving
\beq
I_{1,0}^A=C_{1,0}\xi^{\frac{1}{2(k+2)}}(1-\xi)^{\frac{1}{2(k+2)}}
F(\frac{3}{k+2},\frac{k+3}{k+2},\frac{k+4}{k+2};\xi).
\label{eqn:A10block}
\eeq
These results can be compared with the solutions of the Knizhnik-Zamolodchikov equations \cite{Christe:1987cy}.
The chiral blocks (\ref{eqn:A00block}) and (\ref{eqn:A10block}) are two independent solutions 
which are now related to the Ishibashi states $\lishi A; j=0\vert$ and $\lishi A; j=1\vert$, respectively.
The two point function on the disk is now written as a linear sum of these chiral blocks,
\bea
&&\langle\Phi_{1/2, 1/2}(z_1,\bar z_1)\Phi_{1/2, -1/2}(z_2, \bar z_2)\rangle_{{\rm disk}, \widetilde{A,j}}\nn\\
&&\;\;\;=I^A_{0,0}+\frac{S_{j1}}{S_{j0}}\sqrt{\frac{S_{00}}{S_{01}}}I^A_{1,0}.
\eea

There is another type of two point function for isospin $j=1/2$, but with $m_1=m_2=1/2$.
They are computed similarly, as
\bea
&&\langle\Phi_{1/2, 1/2}(z_1,\bar z_1)\Phi_{1/2, 1/2}(z_2, \bar z_2)\rangle_{{\rm disk}, \widetilde{A,j}}\nn\\
&&\;\;\;=I^A_{0,1}+\frac{S_{j1}}{S_{j0}}\sqrt{\frac{S_{00}}{S_{01}}}I^A_{1,1},
\eea
where the two chiral blocks are
\beq
I_{0,1}^A=C_{0,1}\xi^{\frac{2k+1}{2(k+2)}}(1-\xi)^{\frac{1}{2(k+2)}}
F(\frac{k+3}{k+2},\frac{k+1}{k+2},\frac{2k+2}{k+2};\xi),
\eeq
and
\beq
I_{1,1}^A=C_{1,1}\xi^{\frac{1}{2(k+2)}}(1-\xi)^{\frac{1}{2(k+2)}}
F(\frac{1}{k+2},\frac{3}{k+2},\frac{2}{k+2};\xi).
\eeq
The constants $C_{0,1}$ and $C_{1,1}$ are the three point coupling constants of the bulk primaries,
\bea
&&\Phi_{1/2,1/2}(z,\bar z)\Phi_{1/2,1/2}(0,0)\nn\\
&&\;=C_{0,1}\Phi_{0,1}(0,0)\vert z\vert^{-\frac{3}{k+2}}+C_{1,1}\Phi_{1,1}(0,0)
\vert z\vert^{\frac{1}{k+2}}\cdots.\nn\\
\eea


\subsection{B-branes in $SU(2)_k$ model}


The method discussed above may be applied to the B-branes without difficulty.
Consider $SU(2)_k$ B-branes constructed from A-branes by T-dualizing in the direction of 
$U(1)_k$, i.e. (\ref{eqn:su2BCardy}).
They are characterized by $\Lambda={\rm diag}(-1,1,-1)$.
It is discussed in \cite{Maldacena:2001ky} that these B-branes couple only to primary fields with particular values of $m$.
Let us see how this happens in the Coulomb-gas language.
As the Cardy states $\langle\widetilde{B; \eta, j}\vert$ are linear sums of the states
$\langle B(\alpha, \Lambda)\vert$, we consider the one point amplitude for a primary operator $\Phi_{j,m}(z,\bar z)$,
\bea
&&\lishi B(\alpha, \Lambda)\vert V_{j,m}(z)\bar V^\dag_{j,m}(\bar z)\nn\\
&&\;\;\;\times(\mbox{screening operators})\vert 0,0;\alpha_0\rangle,
\label{eqn:1ptAmpForB}
\eea
where $\Lambda={\rm diag}(-1,1,-1)$.
The holomorphic and antiholomorphic Coulomb-gas charges of the above one point amplitude are respectively,
\bea
&&-\alpha+\alpha_{j,m}\nn\\
&&\;\;\;+(\mbox{holomorphic screening charges}),\\
&&-4\alpha_0\rho-\Lambda^T\cdot\alpha+\alpha^\dag_{j,m}\nn\\
&&\;\;\;+(\mbox{antiholomorphic screening charges}).
\eea
The amplitude (\ref{eqn:1ptAmpForB}) is non-vanishing only when the net charge in each sector 
is neutral.
Examining this condition using the expression of $\alpha_{j,m}$ and $\Lambda={\rm diag}(-1,1,-1)$, 
we find that the neutrality conditions are possible only when $m=0$ up to the reflection and periodic symmetry.
Hence the B-branes couple to primary fields with $m=0$ (up to symmetry) only.
We have used a particular vertex operator representation of the primary field from (\ref{eqn:su2vertex}) here but this property should be independent as all the vertex operator representations are equivalent modulo insertion of the screening operators.

Let us now turn to sample computations of two point functions for the B-branes.
For simplicity we focus on the $j_1=j_2=1/2$ case as in the previous subsection and restrict to $k>1$.
Using the same vertex operator representations (\ref{eqn:2ptvertex}) as A-branes, the chiral blocks are computed similarly, but this time with $\Lambda={\rm diag}(-1,1,-1)$.
The neutrality of Coulomb-gas charges constrains the values of the charges on the boundary and the configurations of the screening operators (i.e., how many of them are in the holomorphic / anti-holomorphic sectors).
In the case of $m_1=1/2$ and $m_2=-1/2$, the neutrality conditions are satisfied in two cases:
when $\alpha=(0,0,0)$ and when $\alpha=(-\sqrt{\frac{2}{k+2}},0,0)$.
The former case corresponds to the intermediate state $\Phi_{0,0}$ and the amplitude,  including one holomorphic screening charge $Q_1$, is computed as
\beq
I_{0,0}^B=C_{0,0}\xi^{-\frac{3}{2(k+2)}}(1-\xi)^{\frac{3k+4}{2k(k+2)}}
F(\frac{1}{k+2},\frac{k+1}{k+2},\frac{k}{k+2};\xi).
\eeq
In the latter case, which corresponds to intermediate $\Phi_{1,0}$, there is one screening charge $\bar Q_1$ in the anti-holomorphic sector and we find the amplitude
\beq
I_{1,0}^B=C_{1,0}\xi^{\frac{1}{2(k+2)}}(1-\xi)^{\frac{3k+4}{2k(k+2)}}
F(\frac{3}{k+2},\frac{k+3}{k+2},\frac{k+4}{k+2};\xi).
\eeq
The constants $C_{0,0}$ and $C_{1,0}$ are the same as those for the A-brane cases.
The two point function for B-type Cardy states are then written as linear combinations of these chiral blocks; using (\ref{eqn:su2BCardy}) we find,
\bea
&&\langle\Phi_{1/2, 1/2}(z_1,\bar z_1)\Phi_{1/2, -1/2}(z_2, \bar z_2)\rangle_{{\rm disk},
{\widetilde{B; j,\eta}}}\nn\\
&&\;\;\;=I_{0,0}^B+\frac{S_{j1}}{S_{j0}}\sqrt{\frac{S_{00}}{S_{01}}}I_{1,0}^B.
\eea
In the off-boundary limit this has the same asymptotic behaviour as the A-brane counterpart.
When $m_1=m_2=1/2$ the neutrality condition of the charges can never be satisfied for the B-branes, marking a sharp contrast to the A-brane case.
This indicates that the two point function for the B-branes identically vanishes,
\beq
\langle\Phi_{1/2, 1/2}(z_1,\bar z_1)
\Phi_{1/2, 1/2}(z_2, \bar z_2)\rangle_{{\rm disk}, \widetilde{B; \eta, j}}=0.
\eeq
This result is reasonable as the fusion of $\Phi_{1/2,1/2}$ and $\Phi_{1/2,1/2}$ generates $m=1$ states which do not couple to B-branes.


\section{Discussion}
In this paper we have discussed free field realization of $SU(2)_k$ and parafermion boundary states based on the Coulomb-gas picture and presented a method to compute correlation functions involving such boundaries.
The formalism naturally includes the so-called B-type boundaries of both $SU(2)_k$ and parafermion models.
The examples of chiral blocks computed for $SU(2)_k$ A-branes are solutions of the Knizhnik-Zamolodchikov equations, and it can be checked that the obtained correlation functions have appropriate clustering properties.
Although we have not given explicit examples in this paper, the technique for finding exact correlation functions in the disk topology also applies to the ${\mathbb Z}_k$ parafermion theory (with the 
primary fields of (\ref{eqn:pfvertex})).
As the $N=2$ supersymmetric minimal models \cite{Frau:1991tf, Ito:1989cm, Ito:1990yy, Kuwahara:1990qw, Kuwahara:1990xy, Ohta:1990qv, Parkhomenko:2003gy} differ from the $SU(2)_k$ WZNW model only by the $U(1)_k$ compactification radius, we expect that the same technique should also apply to these models without much difficulty.

We comment on similarity between the boundary states of the $SU(2)_k$ model and those of the critical 3 state Potts model, which is a spin system having a ${\mathbb Z}_3$ symmetry and is described by a CFT at $c=4/5$ \cite{Fateev:1987vh}.
The Potts model is known to have a $W_3$ symmetry and is diagonal with respect to this $W$-algebra symmetry.
There are six conformally invariant boundary states which also conserve the $W$-symmetry, and they  are identified as three fixed and three mixed boundary conditions of the spin system\cite{Cardy:1989ir}.
In addition this model has two boundary states which conserve the conformal symmetry but break the $W$-symmetry.
These two boundary states are interpreted to represent the free boundary condition and the so-called new boundary condition, which is found in \cite{Affleck:1998nq}.
It has been shown in \cite{Fuchs:1998qn} that these eight states are complete in the sense of \cite{Pradisi:1996yd}.
Free field representations of the Potts model boundary states are constructed and correlation functions involving such boundaries are computed in \cite{Caldeira:2003zz}.
The obtained correlation functions are all consistent with the physics of the spin system.
The free field representation of the Potts model consists of two bosons and in \cite{Caldeira:2003zz} twelve boundary states are constructed, six of them corresponding to $\det\Lambda=1$ and $W$-conserving,  the other six to $\det\Lambda=-1$ and $W$-breaking, where $\Lambda$ is a matrix similar to ours but is $2\times 2$ for the Potts model.
Out of the six $W$-breaking boundaries four of them decouple from the rest of the theory, and only the two -- corresponding to the free and new boundary conditions -- can be seen in the analysis of \cite{Affleck:1998nq,Fuchs:1998qn}.
Clearly this situation is familiar to us;
the $W$-breaking boundaries correspond to B-branes, and apart from some special cases 
("free" and "new" in the Potts model, $m=0$ and its identification images in the $SU(2)_k$ model) the symmetry breaking  boundaries decouple from the rest of the theory.
This of course is not surprising since, as is well known, the 3-state Potts model can be seen as the ${\mathbb Z}_3$ parafermions.

By construction the boundary states we have formulated in this paper give desired overlaps.
Besides, so far as we have checked the Coulomb-gas computation yields reasonable correlators on the disk topology.
There are however a few subtle details in the formulation of boundary states which have not been settled in this paper.
Firstly, we have considered states on the non-chiral Hilbert spaces which are simply direct products of left and right, ${\cal H}\otimes \bar{\cal H}$.
In the standard definition\cite{Cardy:1989ir, Ishibashi:1989kg}, however, the Ishibashi states are built on
asymmetric Hilbert spaces
\beq
{\cal H}\otimes U\bar{\cal H},
\label{eqn:LeftRightHilbert}
\eeq
where $U$ is an antiunitary operator.
This operator $U$ acts non-trivially on the $SU(2)$ currents and as a consequence our expressions of the gluing conditions (\ref{eqn:Jgluing}) differ from some of the standard literature.
If we wish we could have, for example, defined the operator $\Lambda$ (and also $\omega$) as antilinear, rather than linear, and adopted definitions of non-chiral Hilbert spaces which are closer to the traditional one (\ref{eqn:LeftRightHilbert}).
The second subtle point in our formulation is the BRST structure on which the lattice sums are based;
in the three boson formulation of the $SU(2)_k$ model which we have used, so far as we know there seems to be no proof in the literature of the Felder's theorem, that is, triviality of the cohomology space on $F_{r,s}$ except $(r,s)=(l,n)$ that was needed to derive (\ref{eqn:pfchar}).
The technique described in this paper would be generalizable for example to $SU(N)$ cases but for higher $N$ the BRST structure should be more and more complicated.
Thirdly, although it is straightforward to write down integral expressions of correlation functions on the annular topology, i.e. with two boundaries, we have not checked if these expressions satisfy various constraints for consistency.
It is certainly worthwhile investigating in this direction more carefully.
Finally, once the basic Coulomb-gas formulation of D-branes has been established, there are numerous more general directions to explore. 
We hope to come back to these issues in future publications.

\acknowledgments

This work is in part supported by the Magnus Ehrnrooth foundation (S.H.) and the Academy of Finland
(E.K-V.).
S.K. thanks Hiroshi Ishikawa and Satoshi Watamura for helpful conversations, and John Wheater for useful discussions and continuing encouragement.



\end{document}